\documentclass[prl,preprint,preprintnumbers,showpacs,floatfix]{revtex4}
\usepackage[pdftex]{graphicx}
\usepackage{bm}

\begin{document}

\title{
Signal propagation through dense granular systems 
}

\author{L. Kondic}
\affiliation{
Department of Mathematical Sciences and Center for 
Applied Mathematics and Statistics, New Jersey Institute of Technology, 
Newark, NJ  07102
}
\author{O. M. Dybenko}
\affiliation{
Department of Mechanical Engineering, 
New Jersey Institute of Technology, 
Newark, NJ  07102
}
\author{R. P. Behringer}
\affiliation{
Department of Physics and Center for Nonlinear and Complex Systems, Duke
University,  Durham, NC 27708
}
\date{\today}


\begin{abstract}
The manner in which signals propagate through dense granular systems
in both space and time is not well understood.  In order to
learn more about this process, we carry out discrete element
simulations of the system response to excitations where we
control the driving frequency and wavelength independently.
Fourier analysis shows that properties of the signal
depend strongly on the spatial and temporal scales introduced by the
perturbation. The features of the response provide a test-bed
for any continuum theory attempting to predict signal properties. 
We illustrate this connection between micro-scale physics and macro-scale 
behavior by comparing the system response to a simple elastic model 
with damping.  
\end{abstract}
\maketitle

The issue of stress and energy transport through dense granular matter
(DGM) is of significant importance to a number of applications ranging
from detection of land mines to oil exploration.  Due to the
importance of this problem, a significant amount of research has
been carried out in order to understand the basic physical mechanisms
involved.  The major focus of recent work on this issue has been
on how the application of static forces changes the force structure
within a material.  One approach involves considering typically
small size granular samples exposed to time-independent, often
point-like perturbations.  A substantial range of models has been
proposed, including diffusive~\cite{coppersmith96},
wave-like~\cite{bouchaud94b,socolar_pre03,blumenfeld_04} or elastic
response~\cite{goldhirsch_nature05}. 
More traditional continuum models~\cite{nedderman92,goddard90}) 
commonly assume an elastic or elasto-plastic response.  These 
various models are fundamentally at odds with each other, 
since the basic (possibly continuum limit) equations for these 
different descriptions are of parabolic, hyperbolic, or elliptic nature
with respect to their spatial variables.  Some progress in connecting
discrete and continuum descriptions has been reached by realizing that
the system's response may change depending on the scale or the
state of the system: one can see wave-like response on short (meso)
scales, but elastic response on longer ones~\cite{goldhirsch_nature05}.  
For dense systems, theory and experiment~\cite{goldhirsch_nature05,geng_physicad03} 
suggest that an elastic description is best, but near jamming threshold,
a hyperbolic description may apply~\cite{blumenfeld_04}. 

Understanding signal (such as a compression wave) propagation through
a granular system builds on the approaches used for studies of the
static force response but adds the additional feature of
time-dependence.  Within continuum theory, the issue of `sound'
propagation is often considered via an effective medium
approach~\cite{goddard90,sheng73}.  A typical model in this class
explores the response to a spatially independent perturbation which is
large compared to a particle size, such as the response to a moving
piston.  Other work has used experiments and simulations to
explore the pressure dependence of the sounds speed, and the role
of microsctructure including force chains on signal
propagation~\cite{makse_prl99,jia_prl04,somfai_pre05,hostler_pre05_1}.
Very different issues arise for 1D particle chains, which are
characterized by nonlinear high-order wave-like continuum
models~\cite{nesterenko01}.  An extension of these results to higher
dimensions and realistic granular media remains to be carried out.

This letter concentrates on the response of DGM to space-time
dependent perturbations, with the goal of gaining further insight into
the mechanisms involved in stress and energy transport.  To the best
of our knowledge, this approach has not been considered so far.
We concentrate on the regime where the imposed frequencies are low,
and the wavelengths are large compared to the particle size.  We then
compare the information extracted from discrete element (DEM)
simulations to expectations based on a simple continuum picture,
that includes elastic and diffusive behavior.

We choose a relatively simple granular geometry in two spatial
dimensions with the granular particles constrained between two rough
walls (up-down) with periodic boundary conditions (left-right).  The
walls' position prescribes the volume fraction, $\nu=0.9$.  The
particles are polydisperse disks, with the radii varying randomly in
the range of $\pm 5\%$ about the mean.  For
simplicity, we put gravity to zero.  The particle-particle and
particle-wall interactions are modeled using a soft-sphere model that
includes damping, dynamic friction and rotational degrees of freedom,
as explained elsewhere (e.g.~\cite{kondic_99}).  As appropriate for 2D
disks considered here, we use linear springs~\cite{johnson89}.  The
walls in the simulations are made of particles that are rigidly
attached, thus creating an impenetrable boundary.  The wall particles
are strongly inelastic and frictional.  More detailed
explorations of the variation of the DEM model parameters, or of
the force model itself (e.g., presence of static friction), are
considered elsewhere~\cite{kb_future}.  We note that our preliminary
results show only weak dependence of the response on the details of
the force model, or on the parameters.

The simulations are prepared by very slow compression until the
required $\nu$ is reached.  After this
initial stage, the system is relaxed, the upper boundary is fixed and
the lower boundary is perturbed as $z(x) = z_0 + A \sin (\omega t) sin
(k x)$, where $A,~\omega = 2 \pi f$, and $k = {2\pi/\lambda}$ are the
amplitude, angular frequency, and the wavenumber of the imposed
perturbation, respectively.

The simulation parameters are as follows.  Particle properties: the
force constant $k_n = 4000~{m g/d}$, where $g$ is the acceleration of
gravity, and $m,d$ are the average mass and diameter of a particle;
the system particles have the coefficient of restitution $e_n = 0.9$
and the coefficient of friction $\mu_s= 0.1$; the values for the
(monodisperse) wall particles are $e_n = 0.1$ and $\mu_s = 0.9$.
System properties: $40,000$ particles, with the $x$ dimension being
$L=250~d$.  One reason for using such a large system is to reduce the
boundary and finite size effects.  In addition, as we will see below,
with the parameters used, some important features of propagation are
only visible in such a large system.  Perturbation properties:
Amplitude $A = 0.6d$; $\lambda\gg d$ and $f\ll {c^*/d}$ (where
$c^*$ is the speed of sound in the solid) are varied.

\begin{figure}
\centerline{\includegraphics[height=2.8in]{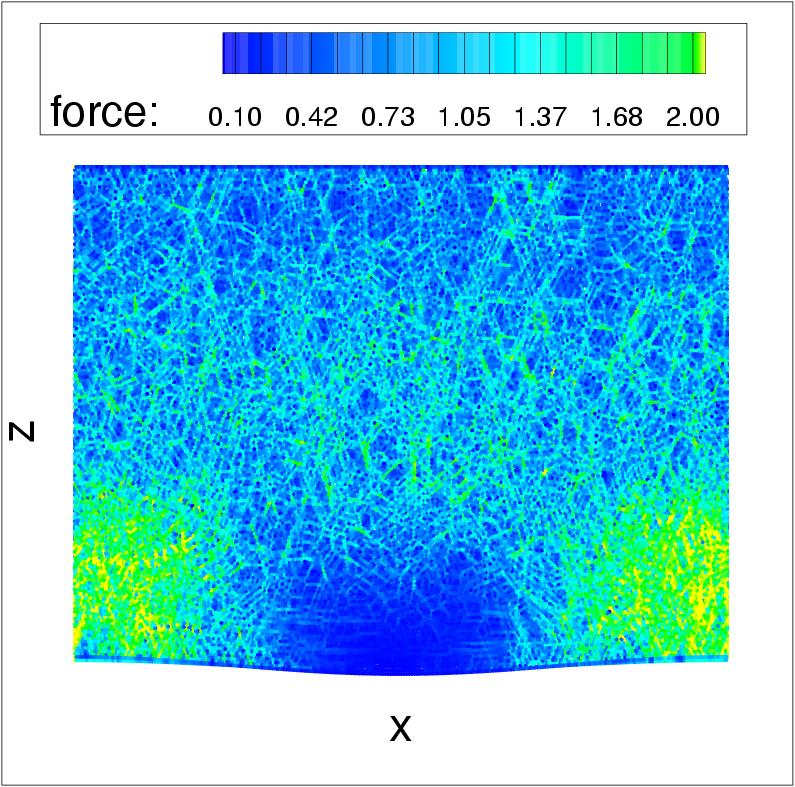}}
\caption{Snapshot of the simulation domain soon after 
perturbation of the lower boundary has been activated.  
The figure shows the forces (normalized by the mean) 
that the particle experience at a given time.
The upper wall is static, and the lower
one performs standing-wave type of motion.  $40,000$ particles.
}
\label{fig:force}
\end{figure}

Figure~\ref{fig:force} shows a snapshot of the simulation domain just after 
the perturbation of the lower boundary has been activated.  The color scheme shows 
the forces on particles, with blue (dark) corresponding to low, and 
green-yellow/light to large forces.
We note the force chains in the interior of the domain, as typically observed
for DGM.

While the evolution, dynamics, and distribution of force chains is of
significant interest in understanding signal propagation in
DGM~\cite{somfai_pre05,hostler_pre05_1}, in what follows we consider
system quantities which can be averaged over a volume which is small
compared to the system size, but which still includes a relatively
large number of particles.  Furthermore, we also carry out time
averaging, choosing an averaging period that is long compared to the
particle collision time, but short compared to $1/f$.  This averaging
procedure is used for calculating the quantities which are later
compared to the results of a simple continuum model.  The presented
results do not depend on the details of the averaging procedure, as
long as the general conditions specified above are followed.

\begin{figure}
\centerline{\includegraphics[height=2.8in]{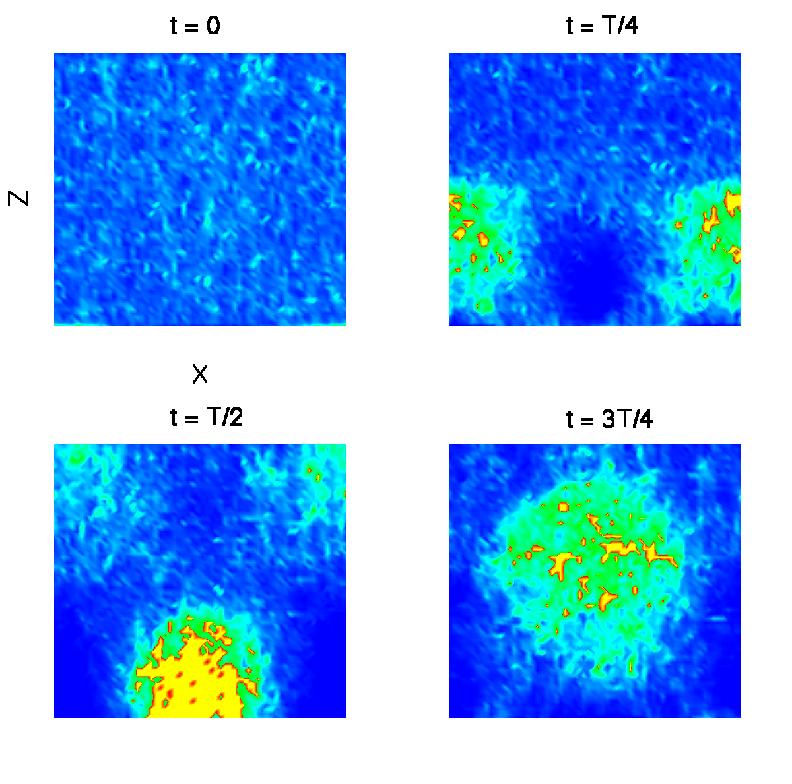}}
\caption{Energy fluctuations for a DEM simulation where a spatially and
temporally harmonic perturbation ($\lambda = 250d$ and $f=30$ Hz) is applied at 
the lower boundary (yellow/light corresponds to high energy and blue/dark to low energy).  
The data in this figure are obtained by subdividing the domain in 
$64$x$64$ cells and than calculating the cell-averages.
}
\label{fig:elas}
\end{figure}

In this work, we concentrate on the elastic energy, which is much
larger than the kinetic one, to illustrate properties of the
propagating signal.  Figure~\ref{fig:elas} shows the elastic energy of
the granular particles at several different phases during one period
of the boundary oscillation~\cite{lk_web}.  We note clear and
well-defined wave forms propagating from the lower towards the upper
boundary.  Analysis of the signal properties such as shown in this
figure is one of the main points of this work.

\begin{figure}
\centerline{\includegraphics[height=2.8in]{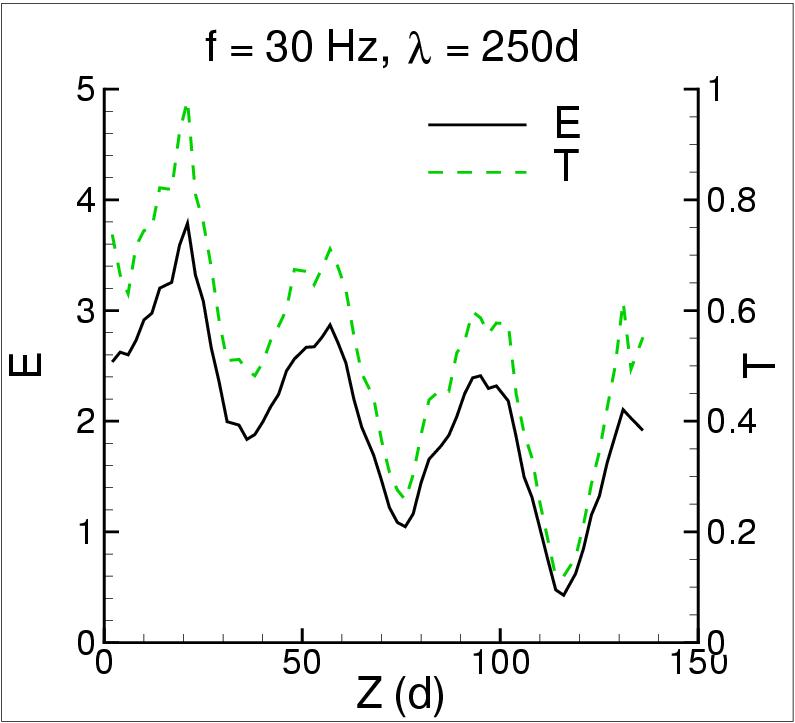}}
\caption{ Dominant Fourier mode (the one imposed by the boundary
  motion) of the elastic energy and temperature as a function of $z$,
  the distance from the oscillating wall.  }
\label{fig:four1}
\end{figure}

Figure~\ref{fig:four1} shows the Fourier transforms (FT's),
$E(z;f,\lambda)$, of the elastic energy, $E$, and of the elastic
temperature defined as $T = <E>^2 - <E^2>$ associated with the
signal~\cite{kb_04}.  We choose here to present the dominant Fourier
mode; the spectral power results are similar.  These FT's are
carried out by expanding the energies accumulated during a given time
interval in the $x$ direction, and then carrying out a time average
over a large number (hundreds or thousands) of periods of the boundary
motion.  Time averaging serves to increase the signal to noise ratio,
and also to ensure that we are not seeing transient effects.  The
initial results, obtained for the first few periods of the
oscillations, are discarded.  We have further verified the steady
nature of the results by carrying out selected simulations for much
longer times.

Figure~\ref{fig:four1} shows a well defined signal propagating in the
$z$-direction.  Clearly, $E$ and $T$, which measures local deviations
of $E$ from the mean, follow each other closely.  Additional
simulations show that the system-selected wavelengths in the
$z$-direction are independent of the system size~\cite{kb_future}.
Next we vary $f$ and $\lambda$ of the perturbation, and discuss how the
spectrum of the resulting signal changes.  We expect that results such
as these will be very useful for comparison with any desired continuum
model.

\begin{figure}
\centerline{\includegraphics[height=2.8in]{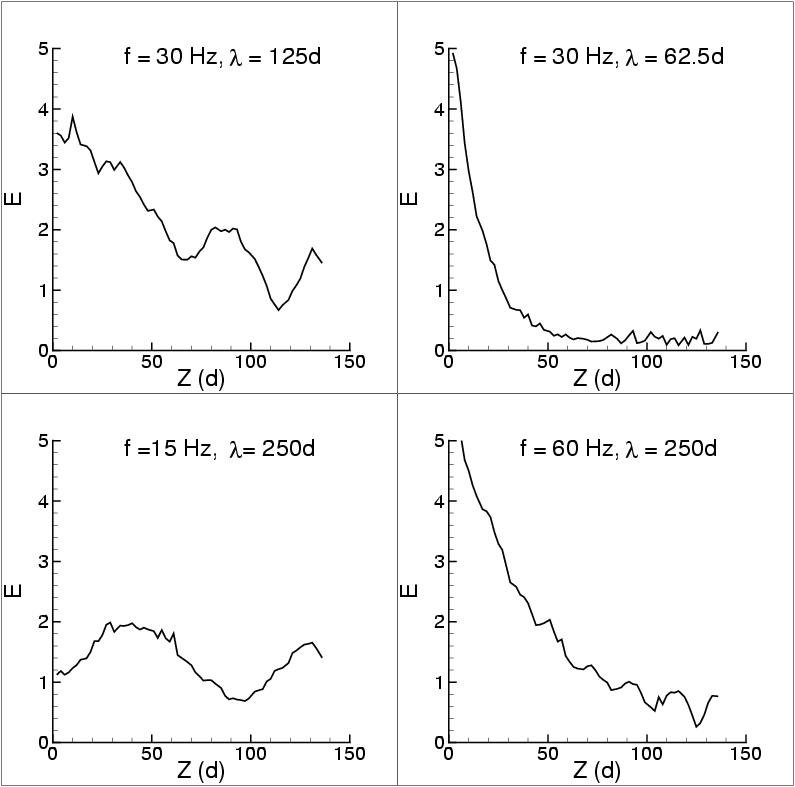}}
\caption{
Dominant Fourier mode of the elastic energy as the frequency and wavelength of 
the perturbation are modified.  
}
\label{fig:four2}
\end{figure}

Figure~\ref{fig:four2} shows {\bf $E(z;f,\lambda)$} as $\lambda$ and
$f$ are varied.  In Figs.~\ref{fig:four2}a-b we see that a decrease of
$\lambda$ leads to a significant dispersion, that is, the wave-like
property of the signal disappears, and the information about the
length-scale introduced by the perturbation tends to be lost.  Note
that although $\lambda$ is decreased, it is still large compared to
the particle size.  Therefore, we are not in the regime where finite
particle size should be important.  Figures~\ref{fig:four2}c-d show
that an increase of $f$ also leads to the loss of the wave-like signal
properties.  For a smaller $f$, we see a signal which is weaker and
also characterized by a larger $z$-direction wavelength compared to
Fig.~\ref{fig:four1}.  Therefore, there is just a narrow regime
  of $f$'s and $\lambda$'s where a well defined signal propagates.
For even smaller $f$'s, the energy input is not sufficiently large to
be traced in an accurate manner.  We note in passing that simulations
with an effectively infinite $\lambda$ were also carried out,
with the results regarding e.g., the speed of propagation, consistent
with the ones in literature~\cite{hostler_pre05_1}.

We next compare these results to a model which includes elastic
behavior and diffusive damping.  We chose such a model since static
force transmission well above jamming densities is, to date, best
described in terms of an elastic
picture~\cite{goldhirsch_nature05,geng_physicad03}, and because we
anticipate presence of dissipative processes.
Assume that the space-time properties of a
considered variable $E$ (but similarly for pressure, temperature, or a
component of the stress tensor) can be described by
\begin{equation}
\nabla^2 E - {1\over c^2} {\partial^2 E \over \partial t^2} - 
{1\over D} {\partial E \over \partial t} = 0\ ,
\label{eq:wave_diff}
\end{equation}
which in the limit $D\rightarrow \infty$ reduces to a linear wave
equation often used to describe wave propagation in elastic solids,
and in the limit $c\rightarrow \infty$ to a diffusion equation, as
considered, e.g. in~\cite{jia_prl04}, although for typically much
larger $f$'s.  We emphasize here that this model in its present form
is not meant as a complete description of the simulation results
presented so far, but instead as a basis for comparison of
signal properties that we understand, and those what we do not.

Eq.~(\ref{eq:wave_diff}) is given in a nondimensional form obtained by
choosing $L$ as a length scale, and $1/\omega_p$ as a time scale,
where $\omega_p$ is some typical imposed frequency (we use $\omega_p =
2\pi f_p$, $f_p = 30$~Hz).  We take the diffusion $D$, and
the speed of propagation, $c$, as constants. 
Further assuming plane wave solution as a zeroth order 
approximation to more complex waves that can be expected
in DGM~\cite{nesterenko01,hostler_pre05_1,sen_granmatt01} in 
the form $E(x,z,t) = E_0 e^{\imath \omega t} e^{\imath k x} e^{\imath q z} $, one obtains
the following dispersion relation 
\begin{equation}
q = - |q| e^{\imath {\phi/2}},\quad |q^2| = {\cal X}^2 + ({\omega/D})^2, \quad
\tan \phi = - {\omega\over (D {\cal X})} \ , 
\label{eq:disp}
\end{equation}
where ${\cal X} = {(\omega/c)}^2 - k^2$.  We first discuss general
features of such a mode in light of the simulation results, and
note that the results outlined below apply for a wide range of
(constant) $c$'s and $D$'s.  The following predictions are easily
verified: (i) for a fixed frequency of perturbation $f$, an increase
of $k$ leads to an increase of the dominant wavelength of the
propagating signal, in agreement with Figs.~\ref{fig:four1} and
\ref{fig:four2}a; (ii) still for a fixed $f$, an increase of $k$ also
leads to larger $Im(q)$ showing that stronger attenuation is expected
for shorter $\lambda$'s, in agreement with
Figs.~\ref{fig:four1},~\ref{fig:four2}a and~\ref{fig:four2}b; (iii)
for a fixed $k$, as $f$ is decreased, one expects longer emerging
wavelengths, in agreement with Fig.~\ref{fig:four2}c.  We note here
that, while the attenuation of a propagating signal as a function of
driving frequency has been considered
before~\cite{jia_prl04,hostler_pre05_1}, we are not aware of any
results discussing the dependence of attenuation on the spatial scales
introduced by the perturbation.  One feature of the DEM results which
is not explained well by the model is the fact that
Eq.~(\ref{eq:wave_diff}) predicts essentially constant attenuation for
larger $f$'s, while in Fig.~\ref{fig:four2}d we see stronger
attenuation than e.g. in Fig.~\ref{fig:four1}, consistent with the
previous work~\cite{jia_prl04,hostler_pre05_1}.  One explanation for
this difference is that the model predicts shorter emerging
wavelengths for these high frequencies. These shorter wavelengths may
become comparable to the particle size, where a continuum model is not
expected to apply.

After finding reasonable agreement between the model,
Eq.~(\ref{eq:wave_diff}), and the simulations, we next ask whether the
values of $D$ and $c$ deduced by comparison to the DEM data are in
qualitative agreement with the commonly used ones.  For this
purpose, we extract the value of $q$ from Fig.~\ref{fig:four1}, and
using the dispersion relation, Eq.~(\ref{eq:disp}), obtain $c \approx
0.025$, $D\approx 0.01$.  Similar values of $c$ and
$D$ can be extracted from the results shown in Fig.~\ref{fig:four2}
and from additional simulations using other values of $f$ and $\lambda$ (not shown), 
typically with the spread of obtained values of $D$ larger than the one for $c$.  
The value of $c$ can now be compared with the speed of sound resulting from
elasticity theory, $c* = \sqrt{E_y/((1-\sigma^2)\rho)}/(L \omega_p)$,
where $E_y,\sigma,\rho$ are the Young modulus, Poisson ratio and
density of the material.  These material parameters can be extracted 
from the DEM force model~\cite{kondic_99} (the force constant there
corresponds to photoelastic disks such as those used
e.g. in~\cite{geng_physicad03}), giving $c/c*\approx 0.1$, in general agreement
with other works~\cite{somfai_pre05}.

An interpretation of $D$ is more complicated.  An estimate based on
the particle size and some typical shear rate, such as average
velocity gradient, underestimates significantly the predicted value of
$D$, suggesting that a different mechanism is in place.
Alternatively, we recall the estimate $D \approx {v_e l/3}$ where
$v_e$ is the velocity of energy transport, and $l$ is the transport
mean free path, measuring the distance traveled before the direction
of propagation is randomized; for further discussion regarding
applicability of this concept to dense granular materials, 
see, e.g.,~\cite{sheng73,jia_prl04}.  Let us assume
that $v_e\approx c*$.  The value of $D$ predicted by the model then
gives $l$ corresponding to $30-40d$.  It will be of interest to
analyze whether such a long lengthscale may be related to the
correlation length introduced by the force-chain structure, the issue
which has been a subject of considerable
discussion~\cite{liu_prl92,jia_prl04,somfai_pre05,hostler_pre05_1}.
It will be also of interest to explore how $l$ (and therefore $D$)
vary with, e.g., the volume fraction, or the imposed frequency. 
An additional task will be to analyze the influence of
dimensionality of the system, since it has been suggested that in 3D
the correlation length of the force network may be much
shorter~\cite{jia_prl04}.  We note in passing that, in the regime 
considered here, a diffusion model
that was successfully applied to the 3D system driven at high
frequencies~\cite{jia_prl04}, could not 
explain the main features of the DEM
results presented in Figs.~\ref{fig:four1} and~\ref{fig:four2}.

While we find consistent results between the simulations and the
simple continuum model encouraging, we note that much more work is
needed.  Regarding simulations, we have considered propagation for a
given volume fraction, therefore not considering explicitly the
pressure dependence of the speed of propagation.  For smaller volume
fractions (closer to the jamming threshold) the nature of the signal
propagation may be modified, and it remains to be seen how the 
continuum model used here would apply in that case.  Regarding the 
continuum model itself, although we find that it describes well many 
features of the DEM results, better understanding of the attenuation 
properties of the signal is needed.  The degree of agreement with the 
simulations suggests that for more precise comparison one may need to consider 
frequency-dependent $D$.  The question of coupling of different spacial and 
temporal scales needs to be considered as well.  

We hope that probing DGM with both space and time dependent perturbations,
as done here, will be utilized in future theoretical and particularly
experimental efforts to build a more complete picture regarding stress
and energy transport in dense granular matter.

{\bf Acknowledgments} We thank Joe Goddard for useful comments.
This work was supported by NSF Grants No. DMS 0605857 and DMR 0555431.


\end{document}